\documentstyle[12pt,twoside]{article}
\input epsf

\title{\Large\bf QCD SUM RULE ANALYSIS OF LEADING TWIST NON--SINGLET OPERATOR MATRIX ELEMENTS}
\author 
{
\it \bf  N. Chamoun$^1$, G. G. Ross$^1$\\
\small$^1$  Department of Theoretical Physics, University of Oxford,\\
\small 1-4 Keble Road, Oxford, OX1 3NP, United Kingdom. \\
}		
\date{}

\begin{document}
\maketitle
\begin{abstract}
We use QCD sum rules to determine the difference between moments 
of the non--singlet structure functions. This combination decouples
from the singular behaviour of the structure
functions near $x=1$ as calculated in the quark--gluon basis and thus should lead to
improved sum rule predictions. However, we find there are still very
large errors due to higher order 
corrections. In order to refine the error analysis, we study the
effect of renormalon ambiguities on our QCD sum rules results. 
\end{abstract}

\section{Introduction}
The application of QCD sum rules to deep inelastic scattering has
been pioneered by Belyaev and Ioffe \cite{ioffe-dis,ioffe-dis2} and
subsequently developed further by Balitsky et al \cite{bbk}. In
\cite{chamoun} we used QCD sum rules to determine the leading twist-two non-singlet
operator matrix elements (OMEs) contributing to deep inelastic
scattering from nucleon targets and an error analysis was carefully
performed. The comparison with the measured values was disappointing
and we investigated the source of errors in detail. We found large sources of error from both the neglect of higher order terms in the expansion in powers of $1/p^2
$ where p is the momentum flowing 
through the nucleon source (``higher--dimension'' terms) and the
neglect of higher order terms in the 
perturbative expansion in powers of the QCD coupling. The phenomenological analysis is very sensitive to the
singular contributions in the structure functions at $x=1$
corresponding to unphysical quark--gluon kinematics. The authors of
\cite{Bra-Gor-Man} have suggested that this problem can be evaded if
one uses the so--called ``Ioffe--time coordinate space
distributions'', as a suitable alternative to the conventional parton
momentum distributions. Here we examine another way of suppressing the
singular contributions at $x=1$. To do this, we consider specific
combinations of moments which eliminate the effects of the
singular corrections at $x=1$. Unfortunately, there is not a great
improvement in the QCD sum rule predictions. Again there seem to be
anomalously large corrections at $O\left(\alpha_s
\langle\bar{q}q\rangle^2/(p^2)^2\right)$ coming from the radiative
correction to the quark condensate term. We also 
investigate further corrections due to renormalon
ambiguities giving power law corrections which cannot be absorbed in
the terms arising from quark and gluon condensates. Indeed, renormalon singularities and their implications for theoretical
predictions have been studied for some time and it has now become clear that one cannot sensibly
talk of power--suppressed corrections if the ambiguities in the
leading term are not under control \cite{Mueller93,MartSac,sachs}. The
reason the renormalon ambiguity arises is that the QCD sum rule
provides an estimate for an OME where the operator
is normalised at the scale $p^2$. Estimating the OMEs'
dependence on $p^2$ involves perturbative calculations that introduce
the renormalon ambiguities. There is a simple correspondence
between renormalon positions and the power corrections to fixed--order 
perturbative predictions evaluated with an infrared cutoff
\cite{Big-Shi-Ura-Vai}.  Following Webber's approach \cite{webber1}, we
impose such a
cutoff by introducing  a small mass $m_g^2$ into the
gluon propagator and examine the effect on the QCD sum rules
determination of the structure functions moments and whether it is process dependent or not, i.e.~whether the
effect is important for some moments while it is negligible for others.

\section{QCD sum rules for the difference of moments}
\label{Moments--difference sum rules}
Our starting point is the result obtained in \cite{ioffe-dis} for the
quark distributions (valid for intermediate values of $x$) :\\
\begin{eqnarray}
\label{start1}
xu_{v}(x,Q^{2}) + M^{2} A^{\bar{\nu}p}(x,Q^{2}) &=&
\frac{M^{6}}{2 \bar{\lambda}^{2}_{N}}e^{m^{2}/M^{2}} L^{-\frac{4}{9}} \left\{ 
4E_{2}\left( \frac{W^{2}}{M^{2}} \right)x(1-x)^{2}(1+8x) \right.\nonumber\\
&&+\frac{b}{M^{4}} \left(
-\frac{4}{27}\frac{1}{x}+\frac{7}{6}-\frac{19}{12}x+\frac{97}{108}x^{2}
\right)E_{0}\left( \frac{W^{2}}{M^{2}} \right) \nonumber \\
&&\left. +\frac{8}{9}\frac{\alpha_{s}}{\pi}\frac{a^{2}}{M^{6}}
\left[
\frac{46}{9}-\frac{38}{9}x-2x^{2}-\frac{2}{9}\frac{x}{1-x}(1+14x)
\right] \right\}
\end{eqnarray}

\begin{eqnarray}
\label{start2}
xd_{v}(x,Q^{2}) + M^{2} A^{\nu p}(x,Q^{2}) 
&=&\frac{M^{6}}{2 \bar{\lambda}^{2}_{N}}e^{m^{2}/M^{2}}
L^{-\frac{4}{9}} \left\{
4E_{2}\left( \frac{W^{2}}{M^{2}} \right)x(1-x)^{2}(1+2x) \right. \nonumber \\
&&+\frac{b}{M^{4}} \left(
-\frac{4}{27}\frac{1}{x}+\frac{7}{6}-\frac{11}{12}x-\frac{7}{54}x^{2}
\right)E_{0}\left( \frac{W^{2}}{M^{2}} \right) \nonumber \\
&&+\frac{16}{9}\frac{\alpha_{s}}{\pi}\frac{a^{2}}{M^{6}} \frac{1}{1-x} 
\left[ x(1+x^{2})\left(\ln\frac{Q^{2}}{M^{2}x^{2}}+C-1 \right)
\right. \nonumber \\
&&
\left. \left. -\frac{8}{9} + \frac{13}{9}x + \frac{247}{36}x^{2}-6x^{3} \right]
\right\}
\end{eqnarray}
where $m$ is the nucleon mass, C is the Euler constant,
\[ a = -(2\pi)^{2}\langle 0 | \bar{\psi} \psi | 0 \rangle \]
\[ b = (2\pi)^{2} \langle 0 | 	\frac{{\alpha}_ {s}}{\pi} {G^a}_{\mu\nu}{G^a}_{\mu \nu} 
	| 0 \rangle  \] 
\( E_{0}(z) = 1 -e^{-z} \), \( E_{2}(z) =1 -e^{-z}(1+z+\frac{1}{2}z^2)
\),
$W$ is the continuum threshold while \( L=\ln(M/\Lambda)/\ln(\mu/\Lambda) \)
takes account of the anomalous dimension of
the currents. The dependence of $\bar{\lambda}^{2}_{N}$ on the Borel parameter $M^2$ 
is expressed via 
 the mass sum rule \cite{mass}\footnote{In \cite{mass}, the sign of
the anomalous dimension of baryonic currents is wrong resulting in a
wrong expression for $\bar{\lambda}^{2}_{N}(M^2)$ in
\cite{chamoun}. Taking account of this fact affects only slightly the
results and conclusions of \cite{chamoun}.}
\begin{eqnarray}
\bar{\lambda}_{N}^2 e^{-\frac{m^2}{M^2}} = M^6 L^{-\frac{4}{9}} E_{2}\left(
\frac{W^{2}}{M^{2}} \right)+\frac{4}{3}a^2 L^{\frac{4}{9}}. 
\end{eqnarray}
In equations (\ref{start1}) and (\ref{start2}), the first term on the
LHS comes from the resonant saturation with a nucleon double pole while the
second parametrises some of the effects of the non--resonant
background corresponding to a single pole term. The
RHS comes from an evaluation of the deep--inelastic scattering process
in perturbative QCD.

In \cite{chamoun}, these relations were used to estimate the quark
distribution moments $M_n^q$ defined by
\begin{eqnarray}
\label{defmomentquark}
M_n^u	&=& \int_{0}^{1}dxx^nu_{v}(x,Q^2) \nonumber\\
M_n^d	&=& \int_{0}^{1}dxx^nd_{v}(x,Q^2)
\end{eqnarray}
which are related to the OMEs
$A^{u,d}_n={\langle}N|\hat{O}^{u,d}_n|N\rangle$ and the Wilson
coefficients $C_n$ in the operator product expansion (OPE) via 
\begin{eqnarray}
\label{moments sum rules}
M_n(Q^2, \alpha_s(Q^2))&=&
\sum_{\tau}A^{\tau}_{n}(\mu_0^2,\alpha_s(\mu_0^2))
C^{(\tau)}_{n}(Q^{2},\mu_0^2,\alpha_s(\mu_0^2)) 
\end{eqnarray}
where the sum runs over the twist $\tau$ of the operators
$\hat{O}_n$ and  
$\mu_0^2$ is the scale at which the operator $\hat{O}_n$ is
renormalised.

A phenomenological study was performed for the OMEs and it revealed a
large discrepancy when comparing with experiment. In \cite{chamoun} we
argued that this discrepancy was due to a breakdown in the
perturbative analysis used in the QCD sum rules. In particular, the 
$O\left(\alpha_sa^2\right)$ radiative corrections give contributions
to the structure functions proportional to $\frac{1}{(1-x)}$ (which
must be regulated at $x=1$) and these terms gave very large
contributions to the moments suggesting that the perturbation series
in $\alpha_s$ does not converge. Further, there are terms proportional
to $\delta(1-x)$ (not displayed in eqs~(\ref{start1})
and~(\ref{start2}) which apply at intermediate $x$) which give very
large higher dimension terms in the moments and suggest the expansion
in powers of $1/p^2$ does not converge either.

One point of view promoted by Belyaev and Ioffe is that the QCD sum
rule predictions should be used only for the structure functions at
intermediate $x$, keeping away from the troublesome singularities at
$x=1$. However we find this unconvincing as such terms come from
unphysical quark and gluon singularities which must be averaged over
when making comparison with data. This averaging procedure will feed the effects of the
singular terms at $x=1$ to intermediate $x$ causing the same
discrepancy with data as is found in the moment relations (which
represent a particular form of averaging). Another possibility that has recently been proposed \cite{Bra-Gor-Man}
is that the problem    
 can be evaded if
one uses the so--called ``Ioffe--time coordinate space
distributions'', as a suitable alternative to the conventional parton
momentum distributions. 

In this paper, we wish to explore a more straightforward way of
eliminating the singularities at $x=1$ for it is easy to do so even in the
conventional picture. In order to achieve this,  
one considers the difference of two consecutive
moments. To see that this eliminates the effect of singularities at $x=1$ we note that the structure function has the form 
(c.f. eqs~(\ref{start1}) and~(\ref{start2})):
\begin{eqnarray}
q(x)&=& f_1(x)\;+\;f_2(x){\delta}(1-x)\;+\;\frac{f_3(x)}{1-x}
\end{eqnarray}
where $f_1$, $f_2$ and $f_3$ are non--singular functions. By
integrating, one obtains the moments:
\begin{eqnarray}
M^n\;=\;\int_0^1\;x^nq(x)dx&=&\int_0^1\;f_1(x)x^ndx\;+\;f_2(1)\;+\;\int_0^1\frac{f_3(x)}{1-x}x^ndx\;.
\end{eqnarray}
As mentioned in \cite{chamoun}, in order to find the
``physical'' moments one needs to regulate the third term:
\begin{eqnarray} 
A_n&=&\int_0^1\;f_1(x)x^ndx\;+\;f_2(1)\;+\;\int_0^1\frac{f_3(x)x^n-f_3(1)}{1-x}dx\;.
\end{eqnarray}
Now we see  that if we subtract $A_n$ from $A_{n+1}$ the effect of
the singularity $\delta(1-x)$ disappears while that of
$\frac{1}{1-x}$ is reduced to an integral of a regular function (the
effect of the regulating term $\frac{f_3(1)}{1-x}$ cancels).
\begin{eqnarray}
\label{Def-Kn}
K_n=
A_{n+1}-A_{n}&=&\int_0^1f_1(x)x^{n+1}dx-\int_0^1f_1(x)x^{n}dx-\int_0^1f_3(x)x^ndx
\end{eqnarray}

\begin{figure}
  \begin{tabular}{cc} 
   \epsfxsize=8.cm
   \epsfysize=8.5cm
   \epsfbox{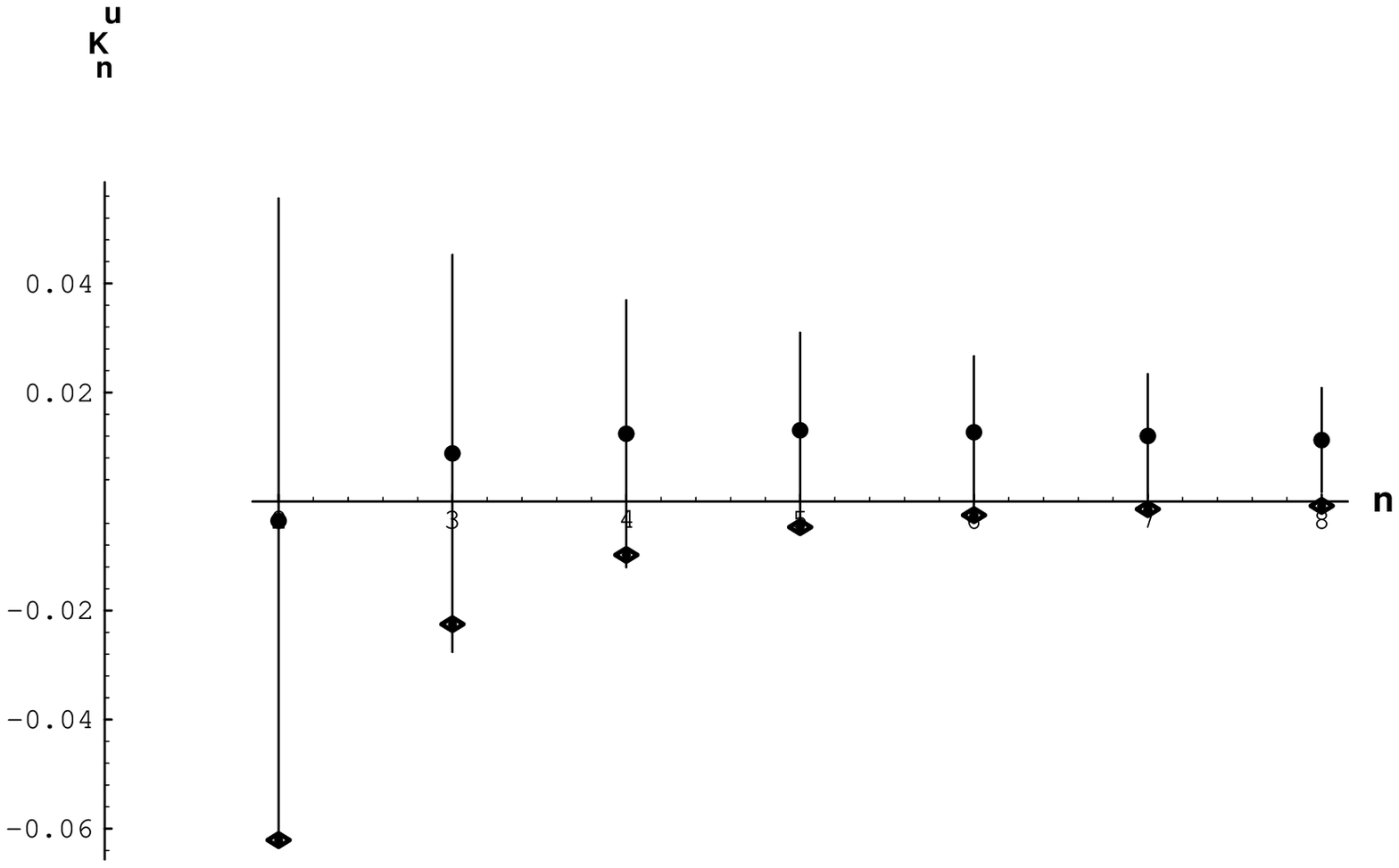 }&
   \epsfxsize=8.cm
   \epsfysize=8.5cm
   \epsfbox{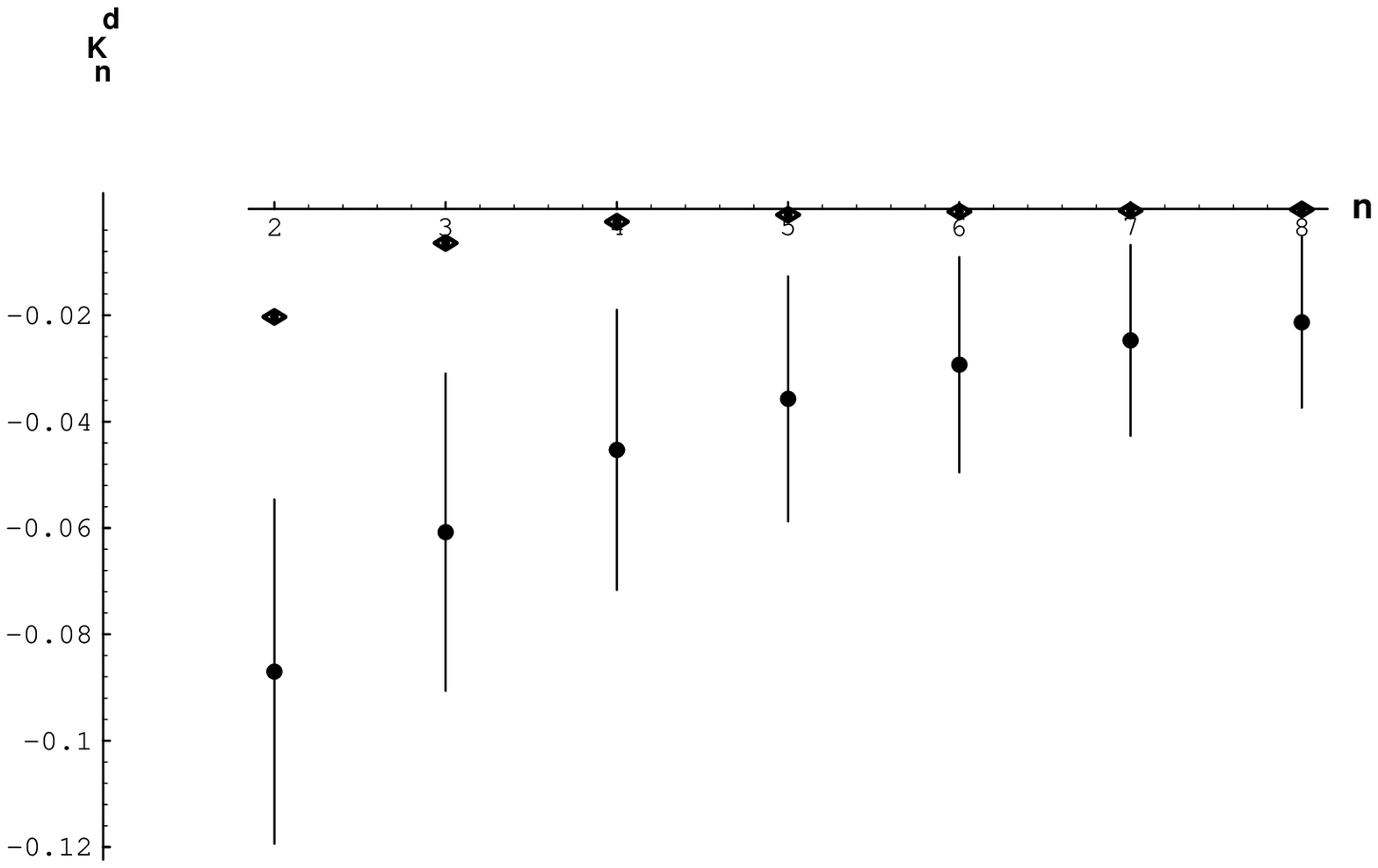 }\\
   (a)&(b)
  \end{tabular}    
    \caption{ Twist--two non--singlet OMEs' difference following from QCD sum rules
together with the error estimates. Also shown are the experimental
determinations of the OMEs' difference (diamonds).}
  \label{figmoments_nosing}
\end{figure}
\begin{figure}
 \begin{tabular}{cc} 
  \epsfxsize=8.cm
  \epsfysize=8.5cm
  \epsfbox{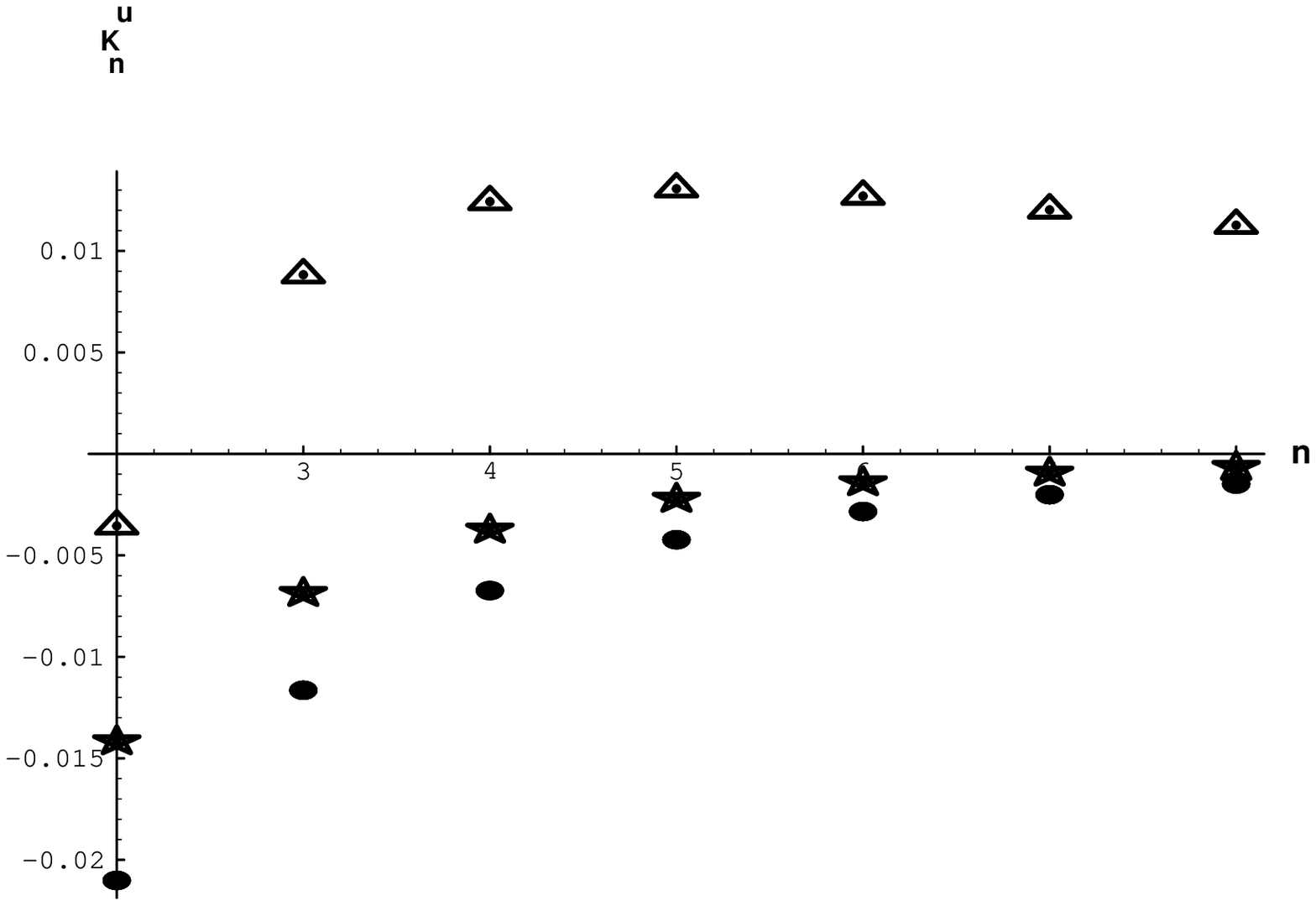}&
  \epsfxsize=8.cm
  \epsfysize=8.5cm
  \epsfbox{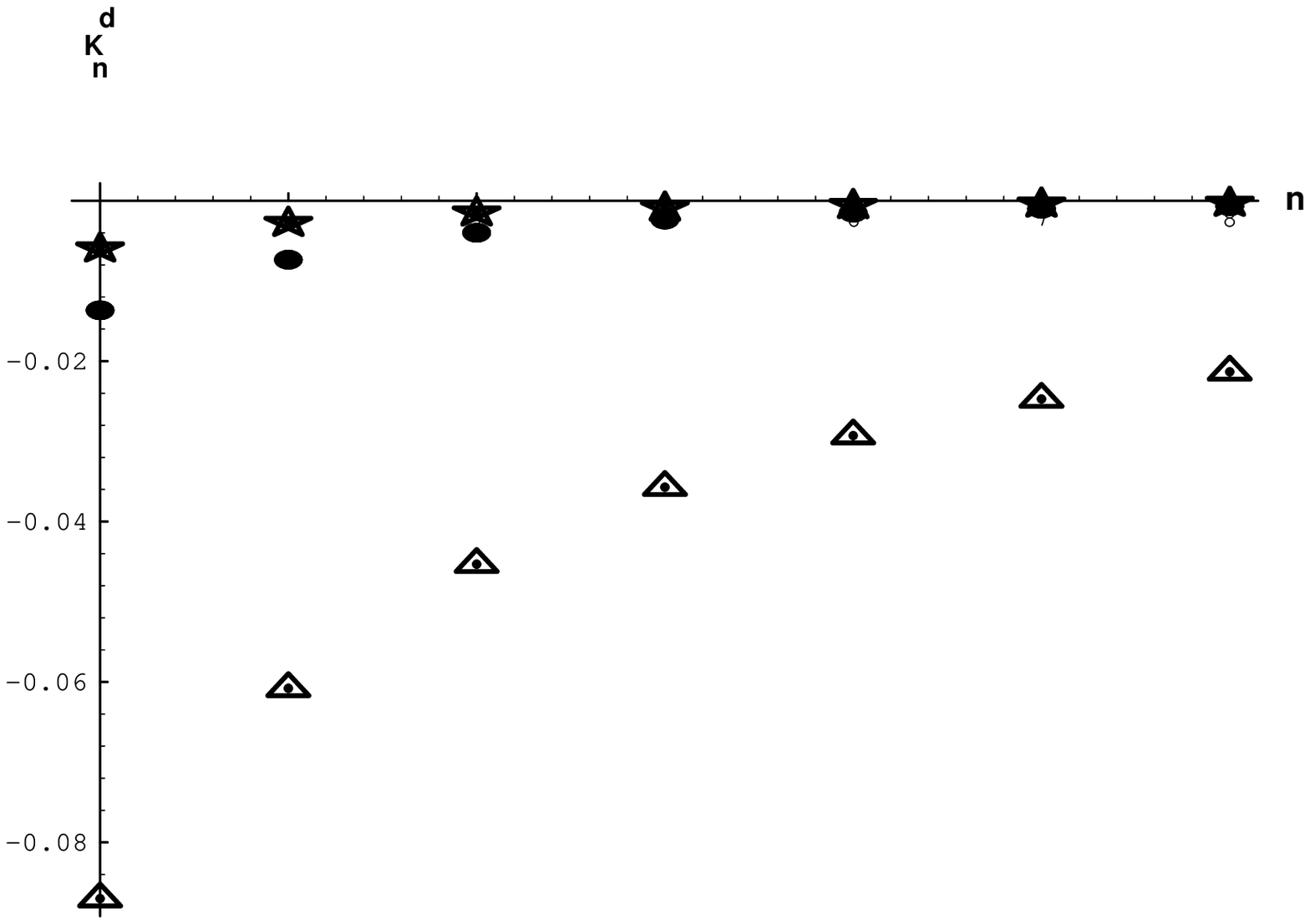 }\\
  (a)&(b)
 \end{tabular}
  \caption{Analysis of the importance of the various contributions to
the QCD sum rules to the OMEs' difference. In (a)  we consider the
up quark operators. The stars show the results neglecting the
gluon condensate and the $O(\alpha_s a^2)$ corrections. The dots
show the results including the gluon condensate. The triangles show the
complete result. (b)  shows the equivalent quantities for the
down quark. }
  \label{succmoment_nosing}
\end{figure} 
Using eqs~(\ref{start1}) and~(\ref{start2}) we may write
down the QCD sum rules for this difference in moments  
and arrive at:

\begin{eqnarray}
\label{Ku}
K_n^u + \frac{M^2}{m^2} D^{u}_n 
&=&\frac{M^{6}}{2 \bar{\lambda}^{2}_{N}}e^{m^{2}/M^{2}}
L^{-\frac{4}{9}} \left\{\right.\nonumber
\\&& 
4E_{2}\left( \frac{W^{2}}{M^{2}} \right)\nonumber\\&&
\left(\frac{1}{n+2}+\frac{6}{n+3}-\frac{15}{n+4}+\frac{8}{n+5}-\frac{1}{n+1}-\frac{6}{n+2}+\frac{15}{n+3}-\frac{8}{n+4}\right)
\nonumber\\
&&+\frac{b}{M^{4}} E_{0}\left( \frac{W^{2}}{M^{2}} \right)\nonumber
\\&&
\!\!\!\!\!\!\!\!\!\!\left(\!\!
-\frac{4}{27}\frac{1}{n} +\frac{7}{6}\frac{1}{n+1}
-\frac{19}{12}\frac{1}{n+2} + \frac{97}{108}\frac{1}{n+3}
+\frac{4}{27}\frac{1}{n-1} - \frac{7}{6}\frac{1}{n}
+\frac{19}{12}\frac{1}{n+1} - \frac{97}{108}\frac{1}{n+2}
\right)\nonumber \\
&&\left. +\frac{8}{9}\frac{\alpha_{s}}{\pi}\frac{a^{2}}{M^{6}}
\left[
\frac{46}{9}\frac{1}{n+1} -\frac{38}{9}\frac{1}{n+2} 
- \frac{2}{n+3} 
-\frac{46}{9}\frac{1}{n} +\frac{38}{9}\frac{1}{n+1} 
+ \frac{2}{n+2}
\right.\right.\nonumber\\&&\;\;\;\;\;\;\;\;\;\;\;\;\;\;\;\;\; \left.\left.
+ \frac{2}{9}\frac{1}{n+1}
+ \frac{28}{9}\frac{1}{n+2}
\right] \right\}
\\
\label{Kd}
K_n^d + \frac{M^2}{m^2} D^{d}_n
&=& \frac{M^{6}}{2 \bar{\lambda}^{2}_{N}}e^{m^{2}/M^{2}}
L^{-\frac{4}{9}} \left\{ \right.\nonumber\\&&
4E_{2}\left( \frac{W^{2}}{M^{2}} \right)
\left(\frac{1}{n+2}-\frac{3}{n+4}+\frac{2}{n+5}-\frac{1}{n+1}+\frac{3}{n+3}-\frac{2}{n+4}\right)  \nonumber\\
&&+\frac{b}{M^{4}} E_{0}\left( \frac{W^{2}}{M^{2}}
\right)\nonumber\\&&
\!\!\!\!\!\!\!\!\!\!\left(\!
-\frac{4}{27}\frac{1}{n}  +\frac{7}{6}\frac{1}{n+1}
-\frac{11}{12}\frac{1}{n+2} - \frac{7}{54}\frac{1}{n+3}
+\frac{4}{27}\frac{1}{n-1} -\frac{7}{6}\frac{1}{n}
+\frac{11}{12}\frac{1}{n+1} + \frac{7}{54}\frac{1}{n+2}
\right) \nonumber \\
&&+\frac{-16}{9}\frac{\alpha_{s}}{\pi}\frac{a^{2}}{M^{6}}
\left[ -2 \left( \psi'(n+2)+\psi'(n+4)\right)
      +2 \left( \psi'(n+1)+\psi'(n+3)\right)\right.\nonumber\\&&
\;\;\;\;\;\;\;\;\;\;\;\;\;\;\;\;\;\;\;\;\;
+\frac{-8}{9}\frac{1}{n}
+ \frac{13}{9}\frac{1}{n+1}
+ \frac{247}{36}\frac{1}{n+2}
+ \frac{-6}{n+3} \nonumber \\
&& \left.\left.
\;\;\;\;\;\;\;\;\;\;\;\;\;\;\;\;\;\;\;\; 
+ \left( \ln\frac{\mu_0^2}{M^2}+C-1 \right) 
\left( \frac{1}{n+1}  + \frac{1}{n+3} \right)
\right] \right\}\;.
\end{eqnarray}
The same phenomenological analysis as in \cite{chamoun} can be done for these sum rules
and the results, for the same numerical values of the
parameters as in \cite{chamoun}, ($a = 0.55 \pm 0.20\,GeV^3$, $b = 0.45
\pm 0.10\,GeV^4$, $\mu=0.5\,GeV$, $\mu_0^2=1\,GeV^2$, $\Lambda = 125
\pm 25\,MeV$, $m=1.00 \pm 0.15\,GeV$ and $W^2=2.3\pm 0.23\,GeV^2$)  are shown in Figs.~\ref{figmoments_nosing}
and~\ref{succmoment_nosing}.

One may see from Fig.~\ref{figmoments_nosing} that there is still a
discrepancy between the prediction of QCD sum rules and the
experimental measurement. To analyse why, we show in
Fig.~\ref{succmoment_nosing} the effect of including various
contributions. One may see that the $O\left(\alpha_sa^2\right)$
contribution is still very large suggesting that the perturbative
series is not convergent.

\section{Renormalon effects}
Given that the QCD sum rule predictions fail to reproduce well either
the moments or the difference of moments it is perhaps appropriate to
consider other possible sources for the discrepancy beyond the immediate
(and depressing) possibility that the expansion in higher dimension
terms and higher order radiative terms is not convergent. We have
identified one possible further error coming from renormalons.

To see how renormalons affect our analysis of the moments, we
look again at the OPE form for the moments, eq~(\ref{moments sum
rules}). There are two scales in our QCD sum rules analysis: $p^2$ and
$Q^2$. The sensitivity to nucleon
momentum  scale $p^2$, which we assume to be large and euclidean in
order to justify the perturbative evaluations of the graphs, resides
in the OME $A_n$ in eq~(\ref{moments sum rules}), while the
dependence on 
the momentum transfer scale, $Q^2$, resides in the coefficient
function $C_n$. By going to large $Q^2$ the coefficient functions may
be reliably calculated, as discussed in \cite{chamoun}. However, the
$p^2$ dependence cannot be reliably calculated in perturbation theory
because the matching condition of QCD sum rules requires values of $p^2
{\approx}1\;GeV^2$ where non--perturbative corrections are
important. Once the $Q^2$--dependence is removed using the OPE the
only scale other than  $p^2$ is $\mu^2$,
the operator renormalisation scale. The latter is in principle
arbitrary, the operator $\mu^2$--dependence being cancelled by the
$\mu^2$--dependence of the coefficient functions. In practice,
however, in evaluating the OME by sum rules, the normalisation scale
must be chosen of the same order as the only other scale $p^2$ to
avoid the appearance of large radiative corrections (which have not
been calculated and thus do not show up explicitly in the analysis). Thus, one should
interpret the result of the QCD sum rule predictions as being for the
operator renormalised at a scale $p^2$. Then, in order to calculate
the OME at a fixed scale $\mu_0^2$ appropriate to the comparison with
the matrix elements extracted from deep inelastic scattering
processes, one must determine the $\mu^2$--dependence of the OME
$A_n(\mu^2)$    
\begin{eqnarray} 
\langle\,N(p)|\hat{O}_{\beta\alpha_{1}{\ldots}\alpha_{n}}(\mu^2)
|N(p)\,\rangle &=& A_n(\mu^2)\;p^{\beta}p^{\alpha_1}{\ldots}p^{\alpha_n}\;+\ldots\,.
\end{eqnarray}
to allow $A_n(p^2)$ to be related to $A_n(\mu_0^2)$.

\begin{figure}
  \epsfxsize=8.cm
  \epsfysize=3.5cm
  \centerline{\epsfbox{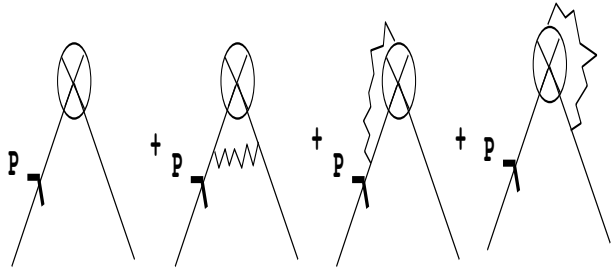 }}
  \caption{Diagrams involved in the calculation of the OME. The solid lines represent quarks, the wavy lines represent gluons
and the circle represents the operator.}
  \label {OME-redressing}
\end{figure}

To estimate
the $\mu^2$--dependence of the OME, one should evaluate the graphs in Fig.~\ref{OME-redressing}.  
\begin{figure}
  \epsfxsize=8.cm
  \epsfysize=3.5cm
  \centerline{\epsfbox{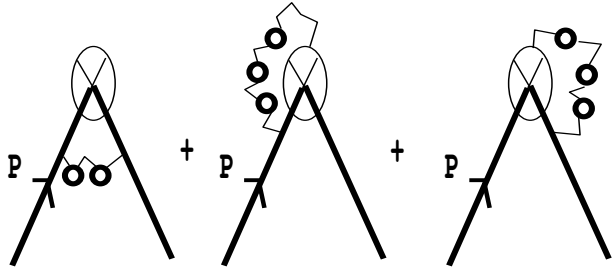 }}
  \caption{Diagrams leading to renormalon ambiguities in $A_n$. The
solid lines represent quarks and the wavy lines represent gluons. The
circles inserted in the gluon lines refer to fermion or gluon loops
and the sum should be over all such insertions.}
  \label {renormalon-redressing}
\end{figure}
The ``renormalon'' contributions are described by graphs such as those in
Fig.~\ref{renormalon-redressing} and in order to take account of their effects, one should introduce
an infrared cutoff in the integrals to be done when evaluating the
graphs. Equivalently one may introduce a mass $m_g$ into the gluon
propagator and expand in powers of $\frac{m^2_g}{-p^2}$, a procedure
which has the advantage of maintaining Lorentz invariance \cite{webber1}.
Using this technique after some calculation one arrives at the following results:

\begin{tabular}{p{1.cm}p{0.1cm}p{14.cm}}
$A_n(p^2)$
&=&
\epsfxsize=8.cm
\epsfysize=2.cm
\epsfbox{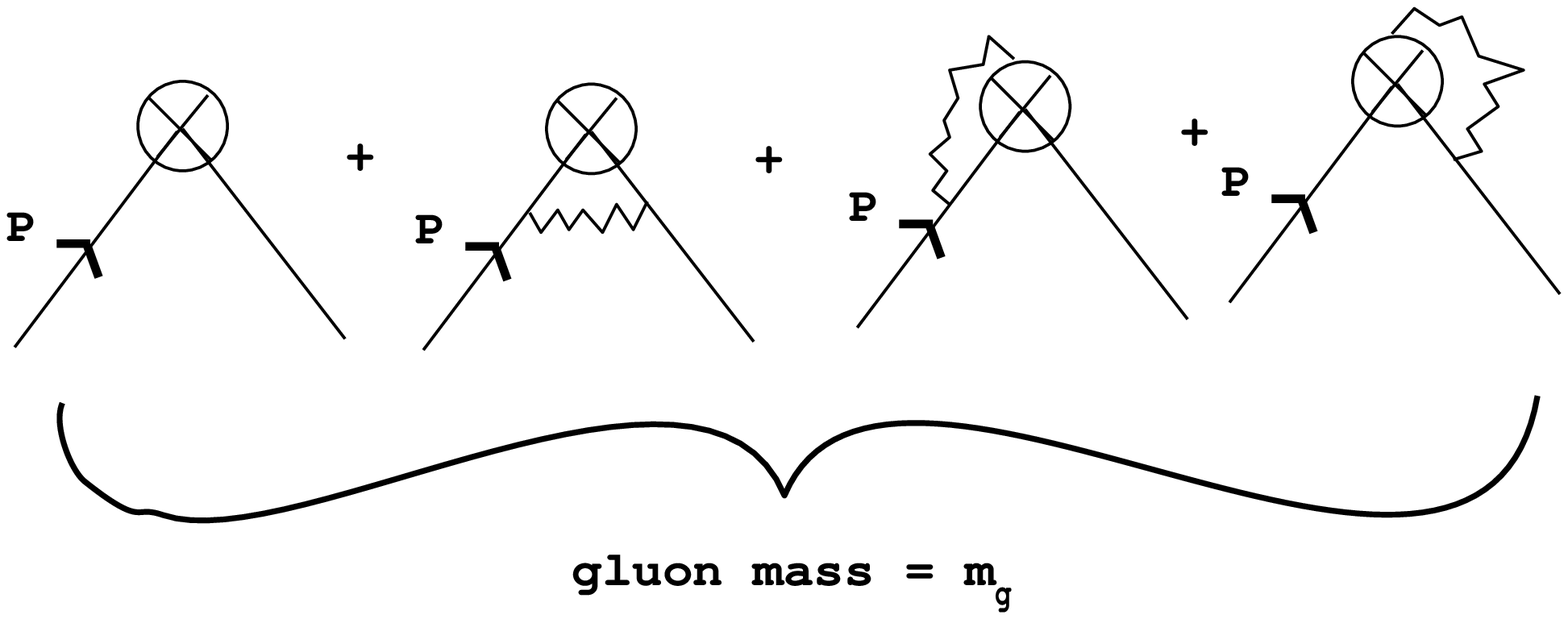}
\\
&=&
$1+\frac{g^2C_F}{8\pi^2}F_1(n)\ln(-p^2)+\frac{g^2C_F}{8\pi^2}F_2(n)
+\frac{g^2C_F}{8\pi^2}F(n)\frac{m_g^2}{-p^2}+\frac{g^2C_F}{8\pi^2}(1-2n) \frac{m_g^2}{-p^2}\ln \frac{m_g^2}{-p^2}
$.
\end{tabular}
\\
where
\begin{eqnarray}
F_1(n)&=&-\left(\frac{1}{n+1}-\frac{1}{n+2}\right)+2\left(\sum_{l=2}^{n+1}\frac{1}{l}\right)\\
F_2(n)&=&\frac{2}{\epsilon}\left(\frac{1}{n+1}-\frac{1}{n+2}\right)+\frac{-4}{\epsilon}\sum_{l=2}^{n+1}\frac{1}{l}+\frac{1}{(n+1)(n+2)}\left(\frac{1}{n+1}+\frac{1}{n+2}+\sum_{l=2}^{n+2}\frac{1}{l}\right)\nonumber\\
&&+\frac{-1}{n+1}+\frac{2}{n+2}-2\sum_{l=2}^{n+1}\frac{1}{l^2}-2n+2\sum_{l=2}^{n+1}\sum_{j=2}^{l}\frac{(-1)^j C^j_l}{lj} \\
F(n)&=&\frac{-1}{n+1}+\sum_{j=1}^{n+1}C_{n+1}^{j}(-1)^{j+1}\frac{1}{j}+2\sum_{l=2}^{n+1}\frac{1}{l}+2\sum_{l=2}^{n+1}\sum_{j=2}^{l}\frac{(-1)^{j+1}C_l^j}{l(j-1)}
\end{eqnarray}
with $C_j^i=\frac{j!}{i!(j-i)!}$ and $C_F$ equals $\frac{4}{3}$ for
$SU(3)$. The singular terms when $\epsilon\rightarrow0$ determine the anomalous dimension of the operator $\hat{O}_n$, while the other terms are finite.

We now introduce the renormalisation scale $\mu^2$ and we get, up to
order $g^2$,
\begin{eqnarray}
\frac{1}{A_n(\mu^2)}&\!\!\!\!=&\!\!\!\!\!1\!-\frac{g^2C_F}{8\pi^2}F_1(n)\ln(-\mu^2)-\frac{g^2C_F}{8\pi^2}F_2(n)
-\frac{g^2C_F}{8\pi^2}F(n)\frac{m_g^2}{-\mu^2}-\frac{g^2C_F}{8\pi^2}(1-2n)
\frac{m_g^2}{-\mu^2}\ln \frac{m_g^2}{-\mu^2},\nonumber\\
\end{eqnarray}
so we can express the OME $A_n$ as
\begin{eqnarray}
\label{OME A_n}
A_n(p^2)&=&A_n(\mu_0^2)\left[1+\frac{g^2C_F}{8\pi^2}F_1(n)\ln\frac{p^2}{\mu_0^2}
+\frac{g^2C_F}{8\pi^2}F(n)\left(\frac{m_g^2}{-p^2}+\frac{m_g^2}{\mu_0^2}\right)\right.\nonumber
\\&&\left.\;\;\;\;\;\;\;\;\;\;\;+\frac{g^2C_F}{8\pi^2}(1-2n)\left(
\frac{m_g^2}{-p^2}\ln\frac{m_g^2}{-p^2}+\frac{m_g^2}{\mu_0^2}\ln \frac{m_g^2}{-\mu_0^2}\right)\right].
\end{eqnarray}
We see from this equation that the relation between matrix elements
for operators renormalised at different scales involves corrections of
$O\left(\frac{1}{p^2}\right)$ corresponding to the mixing of operators
of different twist. This difference gives a further source of error in
the QCD sum rule predictions which we now consider. As we argued
above, one should interpret the QCD sum rule predictions as giving
$A_n(p^2)$ and so the left hand side of the equation has the form
$\frac{A_n(p^2)}{(p^2-m^2)^2}$ for the nucleon pole contribution. 
The
$\ln(\frac{p^2}{\mu_0^2})$ and the $\ln(\frac{m_g^2}{-p^2})$ terms
give, when Borel transformed, terms proportional to
$\ln(\frac{M^2}{-\mu_0^2})$ and $\ln(\frac{M^2}{m_g^2})$
respectively. These terms can be neglected if, for the purpose of
getting a rough estimate of the renormalon effects, we choose $m_g^2 {\simeq}
\left|\mu_0^2\right|{\simeq}1GeV^2$ to be in the QCD sum rules' ``overlapping
window'' ($M^2{\sim}1GeV^2$). For this choice ($m_g^2
{\simeq}-\mu_0^2$), we can also ignore the $\ln \frac{m_g^2}{-\mu_0^2}$
term. That leaves us with the terms proportional to
$\frac{m_g^2}{-p^2}+\frac{m_g^2}{\mu_0^2}$. The former contributes additional
terms corresponding to different poles on the phenomenological side of
the sum rule
\begin{eqnarray} 
\frac{1}{-p^2(p^2-m^2)^2}&=&\frac{\frac{-1}{m^4}}{p^2}+\frac{\frac{-1}{m^2}}{(p^2-m^2)^2}+\frac{\frac{1}{m^4}}{p^2-m^2}.
\end{eqnarray}
The last term corresponds to a single pole contribution so it can be
absorbed by the unknown non--resonant background effect. The first
term is also expected to be small since the structure of eq~(\ref{OME A_n})
applies only for $p^2$ Euclidean and far from $0$. To enforce this we 
introduce a threshold cutoff in the Borel integration of this term,
leading to a vanishing contribution. Finally, the contribution of the
second term is of the same form as that of the term proportional to
$\frac{m_g^2}{\mu_0^2}$ in eq~(\ref{OME A_n}). Combining the two with $-\mu_0^2
{\simeq}+m^2{\simeq}1GeV^2$gives
\begin{eqnarray}
A_n(p^2\!=\!O(\mu_0^2))&\approx&A_n(\mu_0^2)\left(1+2
\frac{\alpha_S(\mu_0^2)C_F}{2\pi}F(n)\frac{m_g^2}{\mu_0^2}\right).
\end{eqnarray}

Physically, we view the gluon mass $m_g^2$ as an infrared matching parameter which
represents the scale below which we switch from the perturbative to
the non--perturbative domain and which is in principle much greater than
$\Lambda_{QCD}$. We shall take it to be a non--perturbative and process--independent
parameter (c.f. \cite{webber1}). If we denote by $A^{QCDSR}_n$ the QCD sum rules predictions of \cite{chamoun} then, to first order in $\alpha_S$, we have
\begin{eqnarray}
\label{A(mu02)-final-expression}
A_n(\mu_0^2)&=&A_n^{QCDSR}\left(1-2\frac{\alpha_S(\mu_0^2)C_F}{2\pi}\,F(n)\frac{m_g^2}{\mu_0^2}\right)\,.
\end{eqnarray}

\section{Discussion}
Eq~(\ref{A(mu02)-final-expression}) shows that renormalon effects
lead to a correction of the QCD sum rule estimate of the OME as
measured in deep inelastic scattering. Following \cite{webber1}, we
shall assume that the $n$--dependence given by the perturbative
calculation with a gluon mass correctly gives the relative magnitude
of the renormalon corrections, the only unknown being the ratio $\frac{m_g^2}{\mu_0^2}$. We
compare this renormalon effect with the deviation $d_n^A$ of the
$A_n^{QCDSR}$ from the experimental data obtained by fitting the
relation
\begin{eqnarray}
\label{A-data}
A^{data}_n(\mu_0^2)&=&A^{QCDSR}_n\left(1+d_n^A\right) 
\end{eqnarray}
between our QCD sum rules predictions $A^{QCDSR}_n$ and our
experimental data $A^{data}_n(\mu_0^2)$ for the OME. The
results are shown in Table~\ref{table-u-quark} for the u--quark and
Table~\ref{table-d-quark} for the d--quark where we have taken the
values $\frac{\alpha_S(\mu_0^2\,=\,-1\;GeV^2)}{2\pi}{\simeq}0.054$,
$\mu_0^2{\simeq}-1\;GeV^2$ and $m_g^2{\simeq}1\;GeV^2$.
\begin{table}
\begin{tabular}{|r||r|r|r|r|r|r|r|r|}
\hline
n 	&   2  	&   3  	&   4  	&   5  	&   6  	&   7  	&   8  	&   9 
\\ \hline
F(n)	&0.5  	&-0.83	&-2.75	&-5.12	&-7.85	&-10.89	&-14.20	&-17.75
\\ \hline
$A^{data}$(n)	&0.11	&0.043	&0.021	&0.011	&0.0064	&0.0039	&0.0025
&0.0017 
\\ \hline
$A^{QCDSR}$(n)	&0.32	&0.31	&0.32	&0.34	&0.35	&0.36	&0.37
&0.38
\\ \hline
$d^A$(n)	&-0.67	&-0.86	&-0.94	&-0.97	&-0.98	&-0.99	&-0.99	&-0.995
\\ \hline
$Ren^A$(n) 	&0.072	&-0.12	&-0.40	&-0.74	&-1.13	&-1.57	&-2.04
&-2.56
\\ \hline
$d^A$(n)/$Ren^A$(n)
	&-9.26	&7.18	&2.36	&1.31	&0.87	&0.63	&0.49	&0.39
\\ \hline
$K^{data}$(n)
	&-0.062	&-0.023	&-0.0098&-0.0047&-0.0025&-0.0014&-0.0008&
\\ \hline
$K^{QCDSR}$(n)	&-0.0035&0.0088	&0.012	&0.013	&0.013	&0.012	&0.011	&
\\ \hline
$d^K$(n)
	&16.49	&-3.54	&-1.78	&-1.35	&-1.19	&-1.11	&-1.07	&
\\ \hline
$Ren^K$(n)
	&17.04	&-10.20	&-9.58	&-11.22	&-13.58	&-16.34	&-19.46	&
\\ \hline
$d^K$(n)/$Ren^K$(n)
	&0.97	&0.35	&0.19	&0.12	&0.088	&0.068	&0.055	&
\\ \hline
\end{tabular}
\caption{
\label{table-u-quark}
Results for the u-quark for the numerical values
$\mu_0^2{\simeq}-1GeV^2$ and
$m_g^2{\simeq}1GeV^2$. $K(n) = A(n+1) - A(n)$; $A^{data}$(n) and
$K^{data}$(n) are the
experimental data while $A^{QCDSR}$(n) and $K^{QCDSR}$(n) represent
the QCD sum rules predictions.
$d^A$(n) represents the deviation between data and QCD sum rules
predictions, 
$Ren^A$(n)$=-2\frac{\alpha_S(\mu_0^2)}{2\pi}C_F\,F(n)\frac{m_g^2}{\mu_0^2}$
represents the ``renormalon'' correction, and
$d^A$(n)/$Ren^A$(n) represents the proportion of these two
quantities. $d^K$(n), $Ren^K$(n) and
$d^K$(n)/$Ren^K$(n) represent the same quantities but for the
difference of moments $K(n)$.}
\end{table}
\begin{table}
\begin{tabular}{|r||r|r|r|r|r|r|r|r|}
\hline
n 	&   2  	&   3  	&   4  	&   5  	&   6  	&   7  	&   8  	&   9 
\\ \hline
$A^{data}$(n)	&0.031	&0.011	&0.0047	&0.0023	&0.0012	&0.0007	&0.0004
&0.0003 
\\ \hline
$A^{QCDSR}$(n)	&0.70	&0.61	&0.55	&0.50	&0.47	&0.44	&0.41
&0.39
\\ \hline
$d^A$(n)	&-0.95	&-0.98	&-0.991	&-0.995	&-0.997	&-0.998	&-0.999	&-0.999
\\ \hline
$Ren^A$(n) 	&0.072	&-0.12	&-0.40	&-0.74	&-1.13	&-1.57	&-2.04
&-2.56
\\ \hline
$d^A$(n)/$Ren^A$(n)
	&-13.25	&8.18	&2.50	&1.35	&0.88	&0.64	&0.49	&0.39
\\ \hline
$K^{data}$(n)
	&-0.020&-0.0064&-0.0024&-0.0011&-0.0005&-0.0003&-0.0001&
\\ \hline
$K^{QCDSR}$(n)	&-0.087&-0.061&-0.045&-0.036&-0.029&-0.025&-0.021&
\\ \hline
$d^K$(n)
	&-0.77	&-0.89	&-0.95	&-0.97	&-0.98	&-0.99	&-0.995 &
\\ \hline
$Ren^K$(n)
	&1.42	&2.37	&3.40	&4.43	&5.44	&6.43	&7.38 &
\\ \hline
$d^K$(n)/$Ren^K$(n)
	&-0.54	&-0.38	&-0.28	&-0.22	&-0.18	&-0.15	&-0.13	&
\\ \hline
\end{tabular}
\caption{
\label{table-d-quark}
The same explanation as for the previous table but for
the d--quark.}
\end{table}

One can do the same analysis for the difference of moments $K_n=A_{n+1}-A_n$, to obtain
\begin{eqnarray}
\label{K(mu02)-final-expression}
K_{n}(\mu_0^2)&=&K^{QCDSR}_n\left(1-2\frac{\alpha_S(\mu_0^2)C_F}{2\pi}\frac{A^{QCDSR}_{n+1}F(n+1)-A^{QCDSR}_{n}F(n)}{A^{QCDSR}_{n+1}-A^{QCDSR}_{n}}\frac{m_g^2}{\mu_0^2}\right)
\end{eqnarray}
\begin{eqnarray}
\label{K-data}
K^{data}_n(\mu_0^2)&=&K^{QCDSR}_n\left(1+d^{K}_n\right)
\end{eqnarray}
where $K_n^{QCDSR}$ denotes the QCD sum rules
predictions for the difference of moments $K_n$,
and $K_n^{data}$ are the corresponding experimental values. These
results are also shown in Tables~\ref{table-u-quark} and~\ref{table-d-quark}.

We see from both these tables
that, for higher moments ($n{\geq}5$), the magnitude of the cutoff--dependent
``renormalon'' contribution for the OME $A_n$ is comparable to, or
greater than, the experimental deviation $d^A_n$. This situation extends even to lower
moments in the case of the difference of moments $K_n$. This means
that renormalon corrections cannot be neglected in the QCD sum rule
evaluation of the higher moments. It is possible to account for some
of the discrepancy between the QCD sum rules and experiment via the
renormalon term through a suitable choice of $m_g^2$. However, at best
this only alleviates the problem slightly. A more cynical view is that
the new corrections identified here are just another source of large higher dimension
corrections to QCD sum rules which render the whole method of doubtful
use in determining the nucleon properties. As we have seen, the
problem persists for the subtracted moments so we cannot blame the
unphysical quark and gluon singularities at $x=1$ for the failure of
the sum rules. The disappointing conclusion of this is that in
determining the leading twist OMEs, where good experimental
measurements are available to check the results, the QCD sum rule method fails largely
because of uncontrollable higher dimension corrections. In our
opinion, this casts doubt on the QCD sum rule predictions for other
nucleon properties such as higher twist matrix elements (where the
troublesome higher order corrections have not been computed).

{\bf Acknowledgement:} We are very grateful to M. Birse for valuable discussions.


\begin{thebibliography}{99}
\bibitem{ioffe-dis} V.M. Belyaev and B.L. Ioffe, Nucl Phys. B310
(1988) 548.
\bibitem{ioffe-dis2} V.M. Belyaev and B.L. Ioffe, Int. J. Mod. Phys. A
6 (1991) 1533.
\bibitem{bbk} I.I. Balitsky, V.M. Braun and A.V. Kolesnichenko,
Phys. Lett. B242 (1990) 245.
\bibitem{chamoun} G.G. Ross and N. Chamoun, Phys. Lett. B380 (1996)
151.
\bibitem{Bra-Gor-Man} V. Braun, P. Gornicki and L. Mankiewicz,
Phys. Rev. D51 (1995) 6036.
\bibitem{Mueller93}A.H. Mueller, Phys. Lett. B308 (1993) 355.
\bibitem{MartSac}G. Martinelli and C.T. Sachrajda, CERN--TH/96--117
(1996) (hep--ph/9605336).
\bibitem{sachs} G. Martinelli and C.T. Sachrajda, preprint
CERN--TH. 7517/94 (1995) (hep--ph/9502352);\\
G. Martinelli, M. Neubert and C.T. Sachrajda, preprint
CERN--TH. 7540/94 (1995) (hep--ph/9504217).
\bibitem{NovShiVaiZak}V.A. Novikov, M.A. Shifman, A.I. Vainshtein and
V.I. Zakharov, Nucl. Phys. B249 (1985) 445.
\bibitem{Big-Shi-Ura-Vai}I.I. Bigi, M.A. Shifman, N.G. Uraltsev and
A.I. Vainshtein, Phys. Rev. D50 (1994) 2234.
\bibitem{webber1}B.R. Webber, Phys. Lett. B339 (1994) 148.
\bibitem{mass}B.L. Ioffe, Nucl. Phys. B188 (1981) 317; Erratum:
Nucl. Phys. B191 (1981) 591.

\end{thebibliography}
\end{document}